\documentclass[conference]{IEEEtran}
\usepackage{cite}
\usepackage{amsmath,amssymb,amsfonts}
\usepackage{algorithmic}
\usepackage{graphicx}
\usepackage{textcomp}
\usepackage{xcolor}
\usepackage{balance}
\usepackage[hidelinks]{hyperref}
\usepackage{todonotes}
\usepackage{booktabs}
\usepackage{siunitx}
\usepackage{multirow}
\usepackage{subcaption}
\usepackage{tcolorbox}
\usepackage{xurl}

\def\BibTeX{{\rm B\kern-.05em{\sc i\kern-.025em b}\kern-.08em
    T\kern-.1667em\lower.7ex\hbox{E}\kern-.125emX}}
\begin{document}

\title{
Analysis and Visualization of the Linux Kernel's Software Evolution Using the City Metaphor
\thanks{}
}

\makeatletter
\newcommand{\linebreakand}{%
  \end{@IEEEauthorhalign}
  \hfill\mbox{}\par
  \mbox{}\hfill\begin{@IEEEauthorhalign}
}
\makeatother

\author{\IEEEauthorblockN{Malte Hansen}
\IEEEauthorblockA{\textit{Department of Computer Science} \\
\textit{Kiel University}\\
Kiel, Germany \\
malte.hansen@email.uni-kiel.de}
\and
\IEEEauthorblockN{Lukas Damerau}
\IEEEauthorblockA{\textit{Department of Computer Science} \\
\textit{Kiel University}\\
Kiel, Germany \\
lukas.damerau@email.uni-kiel.de}
\and
\IEEEauthorblockN{Moritz Bugla}
\IEEEauthorblockA{\textit{Department of Computer Science} \\
\textit{Kiel University}\\
Kiel, Germany \\
stu237520@mail.uni-kiel.de}
\linebreakand
\IEEEauthorblockN{David Moreno-Lumbreras}
\IEEEauthorblockA{\textit{Escuela de Ingeniería de Fuenlabrada} \\
\textit{Universidad Rey Juan Carlos }\\
Fuenlabrada, Spain \\
david.morenolu@urjc.es}
\and
\IEEEauthorblockN{Wilhelm Hasselbring}
\IEEEauthorblockA{\textit{Department of Computer Science} \\
\textit{Kiel University}\\
Kiel, Germany \\
        hasselbring@email.uni-kiel.de}
}

\maketitle

\begin{abstract}
The Linux kernel is one of the largest and longest-maintained open source projects in existence.
With more than 40 million lines of code, understanding the kernel's internal structure and assessing its software evolution is a great challenge.
In this paper, we present an approach to visualize the Linux kernel  using our software visualization tool ExplorViz.
We analyze commits from the Linux Git repository using a custom analysis service.
The web-based frontend utilizes the 3D city metaphor for visualization of the software structure.
Metrics such as the number of lines are collected for each file and are accumulated for directories and commits.
Calculated metrics can be mapped to the building's dimensions or be displayed via a heat map.
A wide range of visualization, search, and filter options enable the interactive exploration of the visualized data.
We present both a visualization of all files in the kernel repository and a visual analysis of the code evolution of Rust files in the Linux kernel.

Demo video: https://youtu.be/3ODnERqD8g0

\end{abstract}


\section{Introduction}\label{sec:introduction}
The Linux kernel is among the largest and longest-maintained open-source projects ever developed.
With tens of thousands of source code files, comprehending its internal structure and code evolution presents a significant challenge.
Several tools have been developed and evaluated in research to visualize different aspects of the Linux kernel~\cite{occhipinti2023syn,linuxvis2025}.
The city metaphor is a common approach that has proven effective for visualizing the structure and code evolution of software systems~\cite{codecity1999, wettel2008}.

In this paper, we present a comprehensive visualization of the Linux kernel repository with our web-based software visualization tool ExplorViz~\cite{explorviz2017,explorviz2020}.
The analysis of the files and computation of metrics is handled by a dedicated Code Analyzer service that is integrated into the software system of ExplorViz.
This is a distinguishing factor of our approach compared to visualization frameworks such as BabiaXR that also support the city metaphor, but rely on input from external tools~\cite{moreno2021babia}.

We present the tool architecture that enables the analysis of Git repositories without the need for external tools in Section~\ref{sec:architecture}.
To showcase the performance and capabilities of our code analysis, we first perform an analysis and visualization in Section \ref{sec:kernel} that includes all files in the kernel's repository.
This is complemented by a visual analysis of the kernel's Rust files in Section \ref{sec:rust} to assess the progress and state of the Rust for Linux project. 
The paper is concluded with a summary and outlook in Section \ref{sec:conclusion}.

\section{Tool Architecture}\label{sec:architecture}
ExplorViz uses a distributed architecture for the processing of data from different origins.
As presented in a previous challenge paper, one kind of data that is handled are OpenTelemetry-compliant\footnote{\url{https://opentelemetry.io/}} observation traces, such as those converted from ftrace\footnote{\url{https://docs.kernel.org/trace/ftrace.html}} logs~\cite{hansen2025linuxchallenge}.
Compared to this previous dynamic analysis, we focus in the present paper on static program analysis and the advancements of ExplorViz regarding the city metaphor visualization that enable a comprehensive exploration of the Linux kernel.

A backend component called Code Analyzer enables the automated cloning and analysis of Git repositories, e.g. from GitHub.
As an alternative, an already cloned repository can be used for the code analysis.

\begin{figure*}
  \includegraphics[width=\textwidth]{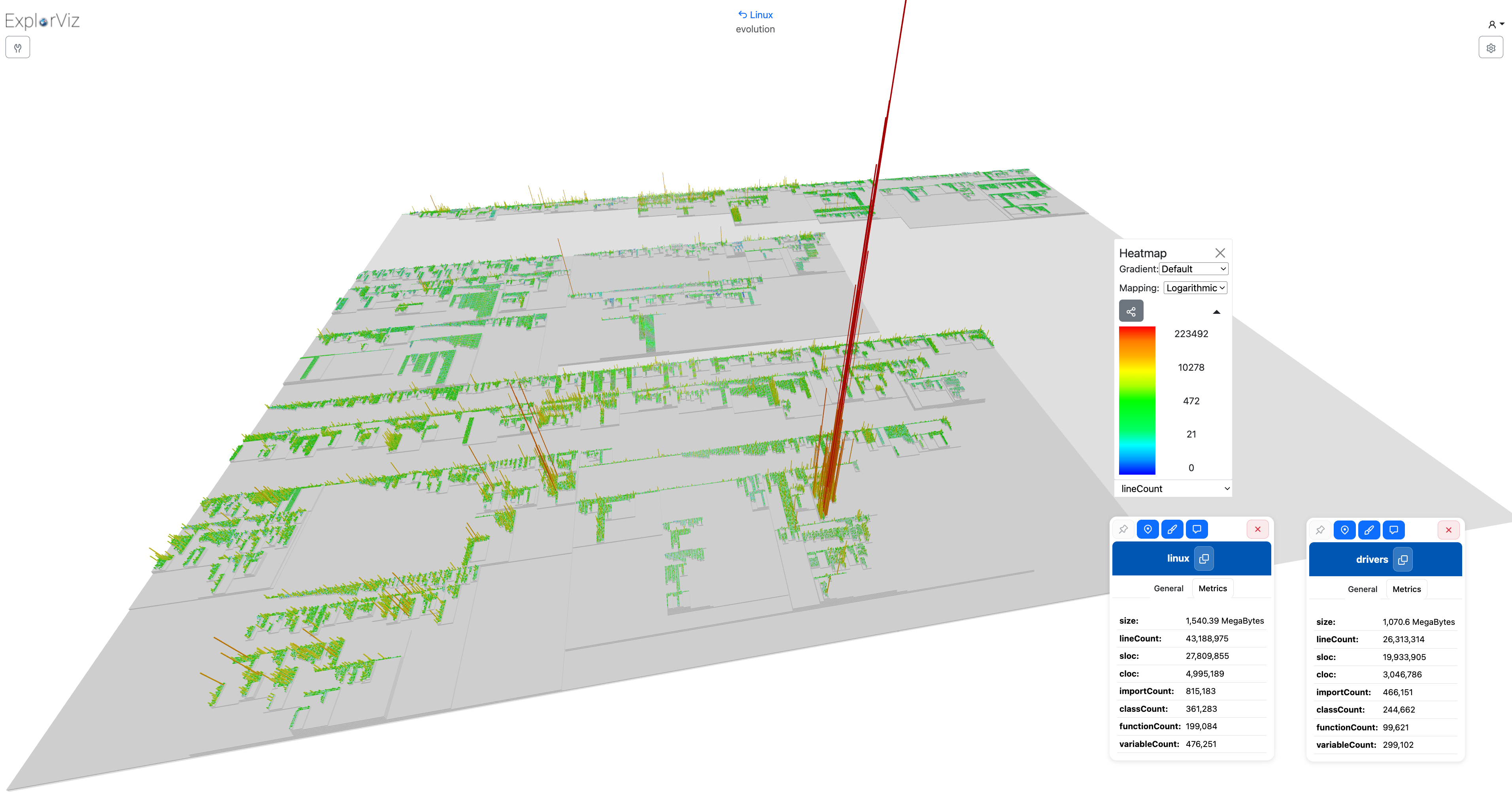}
  \caption{Visualization of all files in the Linux kernel repository in commit \texttt{4539944e5151}.} A heat map with a logarithmic mapping is applied to visualize the line count metric. Two  popups display metrics of the whole Linux repository and the the \texttt{drivers} folder.
  \label{fig:kernel}
\end{figure*}

The Code Analyzer then analyzes each commit in the specified commit range and all corresponding files with respect to the applied filter criteria.
For efficient and customizable static program analysis, we rely on lexers and parsers generated by ANTLR~\cite{parr1995antlr}.\footnote{\url{https://github.com/antlr/grammars-v4}}
All files of the oldest commit in a given commit range need to be analyzed, while only the added and modified files need to be analyzed for subsequent commits.
For each supported source code file, data about its classes and functions is collected.
This excludes the raw source code, as our visualization focuses on the high-level visual exploration of a software system.
The analysis is multi-threaded to reduce the processing time required for repository-scale analysis of the Linux kernel.
For all files, the file size is computed and stored as a metric.
In addition, the following metrics are computed for each supported source code file: lines of code, source lines of code, commented lines of code, number of classes, number of functions, and number of variables.
Among others, the Code Analyzer supports the analysis of C/C++, Python, and Rust source code files, thereby covering the major programming languages that are employed in the kernel.
We do not run an in-depth analysis on the assembly code, shell scripts, and Makefiles that are also present in the kernel.

The resulting data is sent to the Landscape Service, which is responsible for persisting it in a Neo4j graph database.
As directories with their files and commits all exhibit a tree structure, the graph representation of the collected data in a graph database is a straightforward and maintainable technological choice for our approach.

The Landscape Service offers HTTP endpoints such that the commit data and landscape structure (cities, districts, buildings), including a list of metrics for the files, can be requested by the frontend.
More detailed information on individual files are available via additional endpoints on demand such that the size of transferred data structures is kept small.

\begin{figure*}
  \includegraphics[width=\textwidth]{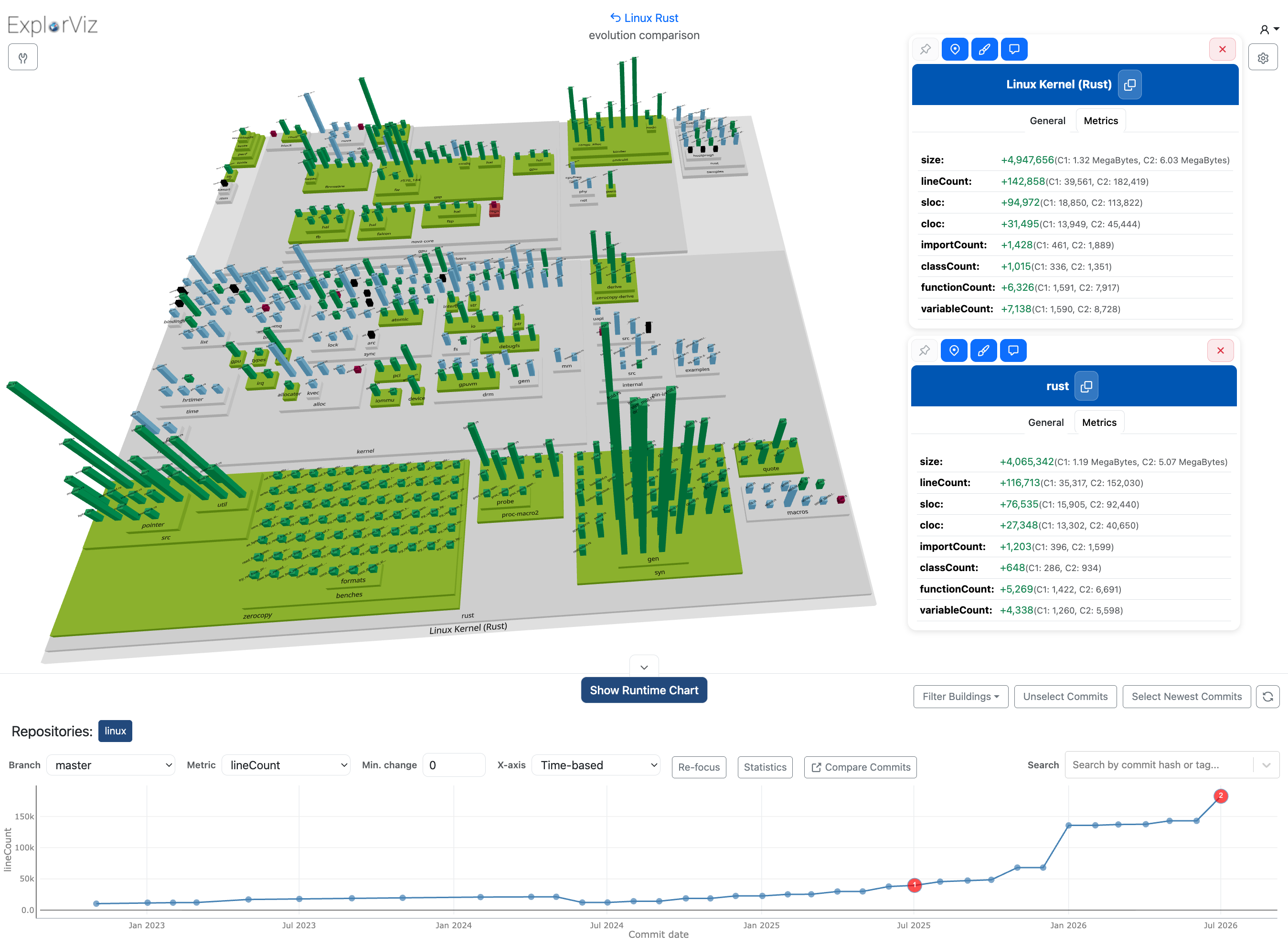}
  \caption{A visual comparison of Rust files in the Linux kernel repository from July 2025 to July 2026. Green elements have been added, red elements have been removed, blue elements indicate modifications, and white/black elements are unchanged. A commit chart in the bottom displays the line count metric on a monthly basis for the analyzed files and highlights selected commits. Two popups display changes in the metrics of the whole repository (on top) and the \texttt{rust} subsystem (on the bottom).}
  \label{fig:evolution-diff}
\end{figure*}

\section{Visualization of the Linux Kernel Repository}\label{sec:kernel}
A user can trigger the code analysis via a web form in the frontend.
When inserting only the GitHub repository's URL and submitting the form, all files in the most recent commit on the default branch are cloned and analyzed.
Thus, the analysis step essentially becomes a one-click action for the user.
After the analysis has been completed, the city metaphor visualization is automatically displayed to the user for the most recent commit.
\autoref{fig:kernel} displays the visualization for a recent commit (\texttt{4539944e5151}) in the kernel's repository.
The city metaphor visualization consists of a city foundation representing the Linux repository, hierarchically stacked city districts representing folders, and buildings representing files.
By default, a color gradient is applied to the districts to indicate the depth of their nesting.
The employed colors can be freely configured in a settings menu.
In the given example, a heat map with a logarithmic mapping is applied to the buildings, indicating the number of lines in the corresponding files.
When a heat map is applied to the buildings, the city foundation and districts are colored white by default such that applied colors can be more easily distinguished.

Regarding the city's layout, a user can choose from a range of layout algorithms provided by the Eclipse Layout Kernel (ELK)~\cite{eclipselayoutkernel2023}.
By default, a rectangle packing algorithm is employed.
However, for the Linux repository with nearly 100,000 files, the computationally more efficient box algorithm\footnote{\url{https://eclipse.dev/elk/reference/algorithms/org-eclipse-elk-box.html}} is employed in our example.
The dimensions of the buildings, i.e. width, depth, and height, can be mapped to the computed metrics.
In the presented example, the height of a building is mapped linearly to the number of source lines of code (SLOC), thereby excluding lines containing only whitespace or comments.

In the visualization example, a cluster of files with a high SLOC and line count metric can be made out.
These files are C header files that reside in the \texttt{driver} directory's \texttt{gpu} subfolder, especially in the \texttt{dcn} folder, i.e. the Display Core Next files for AMD systems.
We also offer a variety of filter options, like the application of regular expressions to narrow down the displayed entities or only display files exceeding given thresholds for metrics.
For example, after applying a filter to display only files with at least 10,000 SLOC, we obtain 149 files (117 C header files and 32 C files) in the visualization.
Further narrowing the search to non-header C files with a large SLOC value, we learn that the largest files are power tables for Realtek devices, e.g. the file \texttt{rtw8852c\_table.c} \footnote{\url{https://github.com/torvalds/linux/blob/4539944e515183668109bdf4d0c3d7d228383d88/drivers/net/wireless/realtek/rtw89/rtw8852c_table.c}} with a SLOC value of 57,117.

\section{Code Evolution of Rust Files}\label{sec:rust}

For decades, the Linux kernel's source code was almost exclusively written in C and assembly language due to their low-level programming support and their efficiency.
However, due to C code's large attack surface, Rust has become more prominent in the kernel~\cite{panther2024rustylinux}.
The Rust for Linux project\footnote{\url{https://rust-for-linux.com/}} has the goal of introducing more Rust code into the Linux kernel, and a RFC to introduce Rust into the Linux kernel was announced in April 2021.\footnote{\url{https://lore.kernel.org/lkml/20210414184604.23473-1-ojeda@kernel.org/}}
Linux kernel Version 6.8, released in March 2024, was the first release with published Rust code, with the nearly three-year gap between the proposal and the release highlighting the complexity of this migration.~\cite{li2024rustforlinux}.

To visually explore the state of Rust in the official mirror of the Linux repository on GitHub, we ran our code analysis on all Rust files for every first commit in a month.
We only included first parent commits on the master branch, thereby excluding commits on feature branches that only became part of the master branch by means of merge commits.
The resulting visualization is presented in \autoref{fig:evolution-diff}.
The commit chart on the bottom displays all analyzed commits where changes in Rust code files have been detected.
The first commit analyzed that includes Rust files was detected in the first November commit of 2022.
This indicates that the first Rust code was included between October 1 and November 1.
The analyzed commit in December in 2022 exhibits no further changes in Rust code, therefore this commit has been skipped in the analysis.

Via a dropdown menu in the commit chart, a metric can be selected that is mapped to the chart's y-axis.
In our example, the line count metric, i.e. lines of code including whitespace and comments, is selected.
It can be seen that, with respect to this metric, the amount of Rust code in the kernel has significantly increased since 2025.
Therefore, we have selected the analyzed commits of July 2025 and July 2026 to explore the recent advancements of Rust code adoption.

The resulting 3D city metaphor visualization in \autoref{fig:evolution-diff} highlights the file-based code evolution.
Green buildings represent added files, red buildings have been removed, blue coloring indicates modifications, and white/black elements are unchanged.
Whenever all files in a folder have been added or deleted, the corresponding district is also colored accordingly.
The height of the buildings is mapped to the SLOC metric.
It can be seen that many files have been added or modified, while only few files have been deleted or stayed unchanged.
The general tab of a city foundation and district popup displays the overall number of contained buildings.
With the ability to filter the visualization to display only unchanged files, it can be determined that only 13 out of 473 files remained unchanged between the selected commits.

For the exemplary visualization in \autoref{fig:evolution-diff}, we have opened popups with metrics of the whole repository and the top-level \texttt{rust} folder.
It can be seen that most of the Rust code resides in this folder and most of the code modifications have taken place in this folder.
Regarding the SLOC metric, only about 0.4\% of the 27.8 million lines in the entire repository are Rust code.
In consideration of these results, it can be assumed that Rust will not completely replace C in the kernel for the forseeable future.

\section{Conclusions and Future Work}\label{sec:conclusion}
In this paper, we presented an approach to perform static program analysis on the Linux kernel repository in the context of its code evolution as well as visualizing the collected data.
We showcased the capability of our approach by analyzing all files of the repository and visualizing them employing the city metaphor.
In addition, we provided a brief visual analysis of the Rust code that has been added to the kernel as part of the Rust for Linux project.
A wide range of filter and configuration options make our approach flexible and enable the exploration of various aspects of the Linux kernel repository.

For future work, we plan to incorporate data on contributor activity, e.g. to identify potential knowledge silos in files that have a small number of contributors.
In addition, we work on the animation of code evolution to make structural and architectural changes in the Linux kernel more visible and easier to comprehend.

\section{Supplementary Materials}\label{sec:material}
The complete software stack of ExplorViz is available on GitHub,\footnote{\url{https://github.com/explorviz}} with complementary documentation available on our website.\footnote{\url{https://explorviz.uni-kiel.de/2-getting-started/}}
Additionally, a demonstration video is available online: \url{https://youtu.be/3ODnERqD8g0}

\providecommand{\doi}[1]{DOI: \href{https://doi.org/#1}{#1}}
\bibliographystyle{myIEEEtran}
\bibliography{bibliography}

\end{document}